%% file: sample-manuscript.tex
\documentclass[sigconf]{acmart}
\copyrightyear{2026}
\acmYear{2026}
\setcopyright{cc}
\setcctype{by}
\acmConference[ICSE '26]{2026 IEEE/ACM 48th International Conference on Software Engineering}{April 12--18, 2026}{Rio de Janeiro, Brazil}
\acmBooktitle{2026 IEEE/ACM 48th International Conference on Software Engineering (ICSE '26), April 12--18, 2026, Rio de Janeiro, Brazil}
\acmPrice{}
\acmDOI{10.1145/3744916.3773111}
\acmISBN{979-8-4007-2025-3/26/04}

\input{setup}

\AtBeginDocument{%
  }
\begin{document}

\title{CREME: Robustness Enhancement of Code LLMs via Layer-Aware Model Editing}

\author{Shuhan Liu}
\orcid{0009-0000-2848-3735}
\affiliation{%
  \institution{Zhejiang University}
  \city{Hangzhou}
  \state{Zhejiang}
  \country{China}}
\email{liushuhan@zju.edu.cn}

\author{Xing Hu}
\orcid{0000-0003-0093-3292}
\authornote{Corresponding authors: Xing Hu and Xin Xia}
\affiliation{
  \institution{Zhejiang University}
  \city{Ningbo}
  \state{Zhejiang}
  \country{China}}
\email{xinghu@zju.edu.cn}

\author{Kerui Huang}
\orcid{0009-0005-7396-5027}
\affiliation{
  \institution{Zhejiang University}
  \city{Hangzhou}
  \state{Zhejiang}
  \country{China}}
\email{huangkerui@zju.edu.cn}

\author{Xiaohu Yang}
\orcid{0000-0003-4111-4189}
\affiliation{%
  \institution{Zhejiang University}
  \city{Hangzhou}
  \state{Zhejiang}
  \country{China}}
\email{yangxh@zju.edu.cn}

\author{David Lo}
\orcid{0000-0002-4367-7201}
\affiliation{
  \institution{Singapore Management University}
  \country{Singapore}}
\email{davidlo@smu.edu.sg}

\author{Xin Xia}
\orcid{0000-0002-6302-3256}
\authornotemark[1]
\affiliation{
  \institution{Zhejiang University}
  \city{Hangzhou}
  \state{Zhejiang}
  \country{China}}
\email{xin.xia@acm.org}

\begin{abstract}
\input{catalogue/abstract}
\end{abstract}



\maketitle
\vspace{-0.2cm}
\section{Introduction}
\label{sec:introduction}
\input{catalogue/introduction}

\vspace{-0.1cm}
\section{Background}
\label{sec:background}
\input{catalogue/background}

\vspace{-0.1cm}
\section{Approach}
\label{sec:methodology}
\input{catalogue/methodology}

\vspace{-0.1cm}
\section{Experimental Setup}
\label{sec:experimentSetUp}
\input{catalogue/experimentsetup}

\vspace{-0.1cm}
\section{Results and Analysis}
\label{sec:results}
\input{catalogue/results}

\vspace{-0.1cm}
\section{Discussion}
\label{sec:discussion}
\input{catalogue/discussion}

\vspace{-0.1cm}
\section{Related Work}
\label{sec:relatedwork}
\input{catalogue/relatedwork}

\vspace{-0.15cm}
\section{Conclusion}
\label{sec:conclusion}
\input{catalogue/conclusion}

\vspace{-0.15cm}
\section*{Acknowledgement}
This research is supported by the National Key R\&D Program of China (No. 2024YFB4506400) and the National Research Foundation, under its Investigatorship Grant (NRF-NRFI08-2022-0002). Any opinions, findings and conclusions or recommendations expressed in this material are those of the author(s) and do not reflect the views of National Research Foundation, Singapore.

\balance
\bibliographystyle{ACM-Reference-Format}
\bibliography{references} 

\end{document}

%% file: setup.tex
\usepackage{graphicx}
\usepackage{multirow}
\usepackage{makecell}
\usepackage{listings}
\usepackage{fancyvrb} 
\usepackage{fontawesome} 
\usepackage{enumitem}
\usepackage{pifont}
\usepackage{graphicx}   
\usepackage{soul}         
\usepackage{xspace}
\usepackage{balance}

\newcommand{\dx}[1]{\textcolor{black}{{#1}}}
\newcommand{\examplebef}[1]{\textcolor{blue!80!black}{\textbf{#1}}}
\newcommand{\exampleaft}[1]{\textcolor{red!80!black}{\textbf{#1}}}

\newcommand{\appname}{{CREME}\xspace}
\newcommand{\appfullname}{\textbf{\underline{C}}odeLLM \textbf{\underline{R}}obustness \textbf{\underline{E}}nhancement via \textbf{\underline{M}}odel \textbf{\underline{E}}diting\xspace }

%% file: catalogue/abstract.tex
Large language models (LLMs) have demonstrated impressive capabilities in code generation, where the natural language prompt plays a crucial role in conveying user intent to the model. However, prior studies have shown that LLMs are highly sensitive to prompt perturbations. Minor modifications in wording, syntax, or formatting can significantly reduce the functional correctness of generated code. As perturbations frequently occur in real-world scenarios, improving the robustness of LLMs to prompt perturbations is essential for ensuring reliable performance in practical code generation.
In this paper, we introduce \textbf{\appname} (\appfullname), a novel approach that enhances LLM robustness through targeted parameter updates. \appname first identifies robustness-sensitive layers by comparing hidden states between an original prompt and its perturbed variant. Then, it performs lightweight parameter editing at the identified layer to reduce performance degradation.
We evaluate \appname on two widely used code generation benchmarks (HumanEval and MBPP) along with their perturbed counterparts. Experimental results show that \appname improves Pass@1 accuracy by 63\% on perturbed prompts while maintaining stable performance on clean inputs, with accuracy deviations within $\pm$1\%. Further analysis reveals that robustness-sensitive layers are primarily concentrated in the middle and deeper layers of the network, and their locations vary across different model architectures. These insights provide a valuable foundation for developing future robustness-oriented editing strategies.

%% file: catalogue/introduction.tex
In recent years, the rapid development of large language models (LLMs) has led to the emergence of powerful models, including ChatGPT~\cite{achiam2023gpt}, LLaMA~\cite{touvron2023llama}, and DeepSeek~\cite{liu2024deepseek}. Trained on large-scale textual corpora, these models exhibit strong generalization capabilities and have achieved notable success across a wide range of software engineering tasks. Among these tasks, code generation has emerged as a key application in AI-assisted software engineering, attracting growing research attention~\cite{chen2021evaluating,li2023starcoder,touvron2023llama,lin2024llm,chen2024code}. 
Typically, practitioners employ LLMs by providing natural language descriptions and LLMs generate the source code, which automates programming tasks and accelerates development workflows.

The natural language description in a prompt is crucial for conveying the requirements defined by users to LLMs. Prior studies~\cite{du2023classeval,chen2021evaluating,austin2021program} have evaluated the code generation capabilities of LLMs using datasets consisting of human-verified prompts. 
However, in real-world scenarios, prompts submitted to LLMs often differ in wording, syntax, and formatting. They may also contain typographical errors or redundant expressions.
Prior studies~\cite{shen2023chatgpt,wang2023robustness,chen2024nlperturbator,shirafuji2023exploring} have shown that LLMs are sensitive to such minor variations, even a slight change may lead to a completely different result. 
Not all users of LLMs are skilled prompt engineers capable of making precise, error-free prompts. Therefore, it is essential to ensure output stability when semantically equivalent prompts contain minor variations. This underscores the importance of improving the robustness of LLMs to natural prompt perturbations.

Existing studies have proposed various strategies to improve the robustness of LLMs, varying from Input-level interventions to model-augmentation approaches.
Input-level interventions~\cite{wang2024resilience,agrawal2025enhancing} aim to sanitize or rephrase perturbed prompts before passing them into the model. For example, LLMs can be used to denoise inputs or generate multiple paraphrased variants, from which the most effective is selected. Although these techniques improve performance under prompt perturbations, they do not modify the internal robustness of the model itself.
In contrast, model-augmentation approaches~\cite{hu2024prompt,agrawal2025enhancing,sun2023evaluating} add trainable components such as soft prompts, retrieval-augmented generation (RAG), or Low-Rank Adaptation (LoRA) to handle input variability. However, these methods increase system complexity and require additional training efforts. 
These limitations underscore the need for a lightweight method to directly enhance the inherent robustness of LLMs without extensive retraining or architectural modifications.

In recent years, knowledge editing techniques have been proposed for LLMs~\cite{li2024model,wang2024knowledge,zhang2024comprehensive}, enabling efficient post-training updates without full model retraining. Therefore, it is an intuitive idea to explore whether such localized modifications can also improve the robustness of CodeLLMs.
However, existing knowledge editing methods mainly tackle factual knowledge~\cite{meng2022locating,li2024pmet,dai2021knowledge,wu2023depn}. They depend on the subject tokens or specific phrases in a single sentence to locate the areas for editing. In contrast, robustness-oriented tasks often involve complex, multi-sentence natural language prompts, making it more difficult to identify meaningful intervention targets.
DINM~\cite{wang2024detoxifying} recently applies knowledge editing to detoxification tasks. While the inputs in these tasks are also complex, the availability of gold-standard safe responses enables direct supervision during editing. In contrast, code generation tasks lack a single correct output, presenting unique challenges for applying knowledge editing in this context.

In this paper, we introduce \textbf{\appname}, a lightweight framework that uses a pair of prompts (i.e., an original prompt and its perturbed variant) to enhance the robustness of LLMs under specific types of prompt perturbations. Unlike traditional robustness enhancement methods, which focus on input modification or additional components, \appname targets the internal mechanisms of the model itself. Specifically, \appname first locates the key layers most responsible for the robustness degradation of the models under perturbations via a layer-wise causal intervention strategy. Then, \appname performs lightweight parameter editing at this layer to align the representations of the perturbed prompt with those of the original prompt, while preserving the behavior of the model on clean inputs.

To assess the effectiveness of the proposed framework, we conduct experiments on two widely used code generation benchmarks: HumanEval~\cite{chen2021evaluating} and MBPP~\cite{austin2021program}, along with their perturbed counterparts provided by NLPerturbator~\cite{chen2024nlperturbator}. These perturbations are designed based on empirical observations of real-world user interactions with code LLMs.
We evaluate our method on two representative open-source LLMs: \texttt{CodeLlama-7b} and \texttt{Qwen2.5-Coder-7B}.
To provide a comprehensive evaluation, we compare \appname with four strong baselines. These include two robustness-enhancement methods (i.e., Self-Denoising~\cite{agrawal2025enhancing} and LoRA Fine-Tuning~\cite{hu2024prompt}) and two knowledge-editing approaches (i.e., ROME~\cite{meng2022locating} and DINM~\cite{wang2024detoxifying}). Experimental results demonstrate that \ding{182} \appname significantly improves model robustness, yielding a 63\% relative increase in Pass@1 accuracy on perturbed prompts. \ding{183} \appname exhibits strong generalization across diverse perturbation types. Editing the model based on a single perturbed instance restores up to 30\% of the overall code generation accuracy within that perturbation category. \ding{184} Causal tracing-based layer localization plays a critical role in robustness enhancement by accurately identifying the robustness-sensitive regions within the model.
\ding{185} Robustness-sensitive layers exhibit a clustering pattern, and their positions shift depending on the model architecture.
\ding{186} \appname maintains stable performance on clean inputs, with accuracy deviations within $\pm$1\%.

\noindent\textbf{Contributions:} In summary, the main contributions of this paper can be summarized as follows:
\vspace{-0.05cm}
\begin{itemize}[leftmargin=*]
    \item We introduce \textbf{\appname}, a novel editing framework that uses a single pair of original and perturbed prompts to identify robustness-sensitive layers and update targeted parameters to enhance the robustness of LLMs. The replication package of our work is publicly available at~\cite{creme2025}.
     \item We propose \textbf{G-RIR}, an evaluation metric designed to quantify the generalization ability of robustness enhancement methods.
    \item  We analyze where robustness-sensitive layers are located within LLM architectures, providing insights to guide future robustness enhancement methods.
\end{itemize}

%% file: catalogue/background.tex
\subsection{Motivating Example}
In real-world scenarios, developers interact with code LLMs by providing natural language prompts to generate desired code completions. However, users may inadvertently introduce minor errors into prompts, such as typographical mistakes, repeated words, or slight phrasing inconsistencies. Although such errors are semantically negligible and easily overlooked by human programmers, they can cause LLMs to produce significantly different outputs.

\begin{figure}[h]
\vspace{-0.1cm}
  \centering  \includegraphics[width=\linewidth]{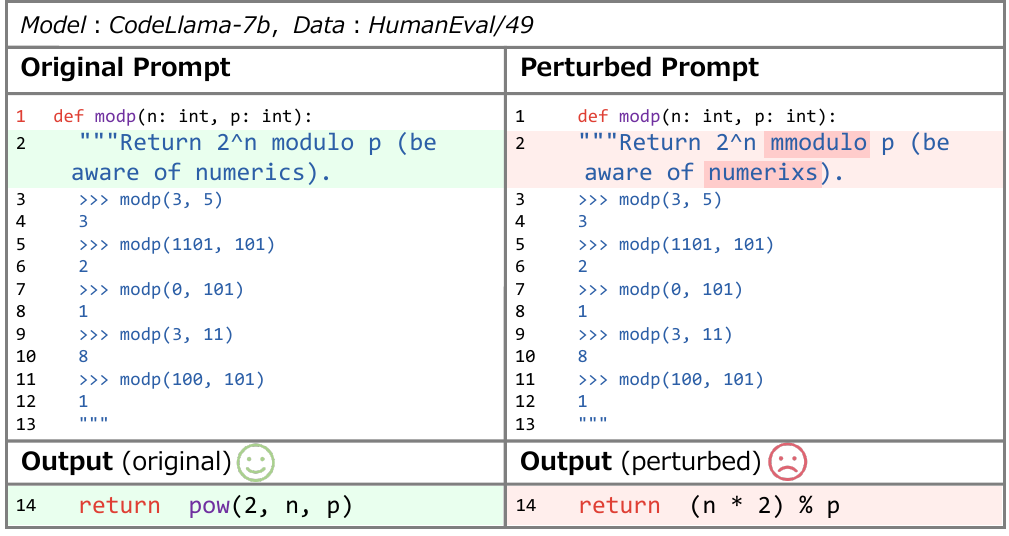}
\captionsetup{skip=2pt}
  \caption{Example of Code Generation using Original and Perturbed Prompts}
    \label{fig:motivation}
\vspace{-0.5cm}
\end{figure}

As shown in Figure~\ref{fig:motivation}, we provide an example using the \texttt{CodeLlama} model to illustrate this issue. The original prompt correctly describes the task, and the model generates a functionally correct implementation that passes the corresponding test cases. 
However, when ``modulo'' is misspelled as ``mmodulo'' and ``numerics'' as ``numerixs'' in this case, the model generates a substantially different implementation that fails the functional tests.

This example highlights a critical limitation of current code LLMs: their generation behavior is sensitive to minor prompt perturbations. In real-world software development scenarios, where prompts naturally vary across users and situations, such weakness reduces the reliability and effectiveness of LLMs. Therefore, there is an urgent need for robustness-oriented techniques that improve model robustness to natural language perturbations without requiring model retraining or explicit prompt rewriting.

\subsection{Task Definition}
We formulate the task of robustness enhancement through model editing in the context of code generation. Let $G$: $P \rightarrow Y$ be a pre-trained autoregressive language model that takes a natural language prompt $P$ as input and generates a code snippet $Y$ as output. Let $P_{ori}$ denote an original, unperturbed prompt, and $P_{pert}$ be a perturbed variant. Although a human programmer would interpret both prompts as describing the same task, the model often produces different outputs:

\vspace{-0.2cm}
\begin{align}
&P_{ori} \approx P_{pert} \quad \text{(slight perturbations)} \quad \not\Rightarrow \quad \notag\\
&G(P_{ori}) \approx G(P_{pert}) \quad \text{(output equivalence)}
\end{align}

To address this issue, we aim to construct a locally updated model $G_{\mathcal{W}'}$ whose behavior on $P_{pert}$ aligns with the robust output for $P_{ori}$, without requiring full retraining or additional external data. Let $\mathcal{W}$ denote the original parameters of $G$. We introduce a robustness editor $\xi$ that modifies only a small subset of $\mathcal{W}$ to obtain the edited model $G_{\mathcal{W}'}$:

\vspace{-0.2cm}
\begin{equation}
G_{\mathcal{W}’} = \xi\left(G_{\mathcal{W}}, (P_{ori}, P_{pert})\right)
\end{equation}

$\mathcal{W}'$ are the edited parameters after applying $\xi$ based on a single prompt pair $(P_{ori}, P_{pert})$. This procedure yields a model that not only produces consistent outputs for $P_{pert}$ but also generalizes to other prompts exhibiting similar types of perturbations, thereby enhancing the robustness of $G$ to natural language variations.

%% file: catalogue/methodology.tex
\begin{figure*}[h]
  \captionsetup{skip=3pt}
  \centering  \includegraphics[width=\linewidth]{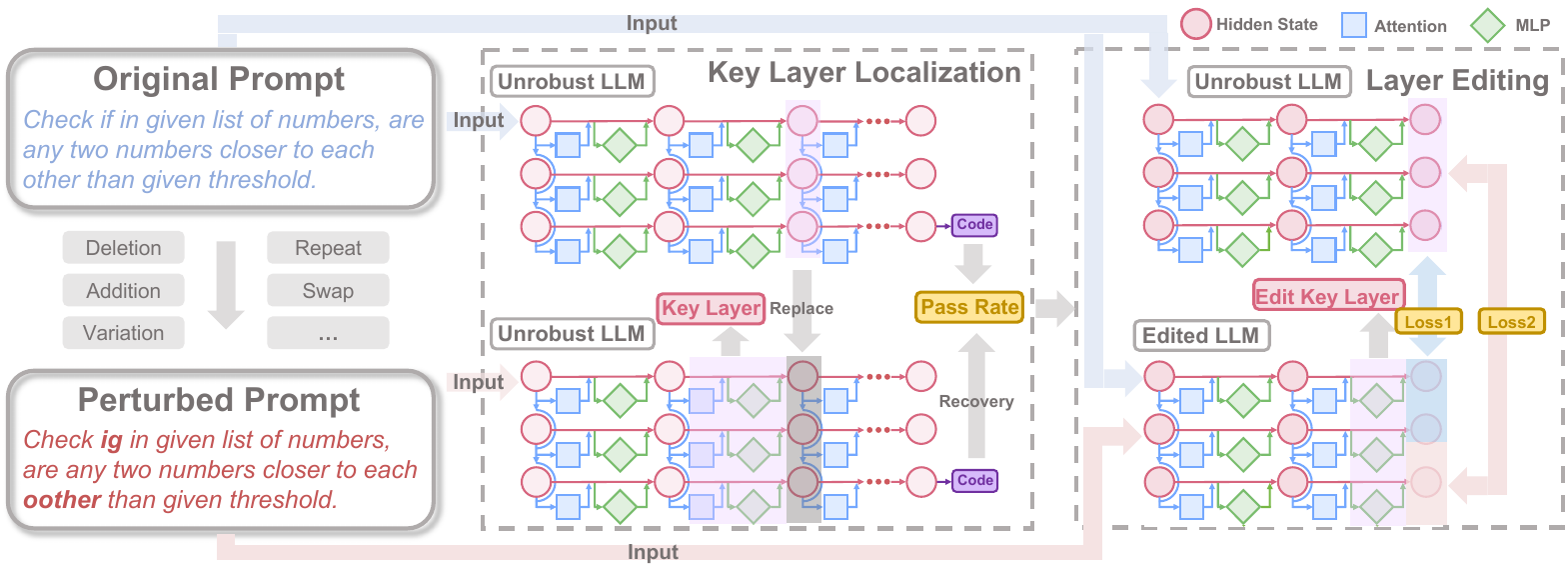}
  \caption{\appname framework: \ding{182} (left) A slight perturbation is inserted to the original prompt. \ding{183} (middle) CREME identifies robustness-sensitive key layers by replacing each layer’s hidden states with those from the original prompt and evaluating recovery in pass rate. \ding{184} (right) The key layer is fine-tuned with two objectives: preservation loss (Loss1), which retains behavior on clean inputs, and alignment loss (Loss2), which enforces consistency between original and perturbed prompts.
}
\label{fig:method}
\vspace{-1em}
\end{figure*}

In this section, we first provide an overview of \appname and its architecture. Then, we introduce each component in detail.

\vspace{-0.1cm}
\subsection{Overview}
In traditional knowledge editing, the primary objective is to identify the key neurons or layers associated with a specific factual statement and modify them to ensure the model internalizes the new knowledge. In contrast, our goal is to enhance the overall robustness of LLMs against perturbed prompts. Given a pair of prompts—an original prompt $P_{ori}$ and its perturbed counterpart $P_{pert}$—we aim to adjust the model such that its generation behavior on $P_{pert}$ closely aligns with that on $P_{ori}$, thereby preserving functional correctness. This differs from traditional knowledge editing in two key aspects: \ding{182} Prompts often have complex expressions, making it difficult to identify a clear subject;
\ding{183} The success criterion is based on the functional correctness of the generated code, rather than matching a fixed output.

To address these challenges, we propose \textbf{\appname} (\appfullname). Notably, our approach requires only a single pair of prompts (i.e., $P_{ori}$ and $P_{pert}$) to improve the model’s robustness against a specific type of prompt perturbation. As illustrated in Figure~\ref{fig:method}, \appname comprises two main components. First, we perform a causal analysis to identify the key layer responsible for robustness degradation. Then, we apply representation-aligned model editing at the identified key layer to mitigate the effects of prompt perturbations. 
Each component is detailed in the following subsections.

\vspace{-0.1cm}
\subsection{Key Layer Localization}
To identify the layer most responsible for robustness degradation under prompt perturbations, we conduct a layer-wise causal intervention procedure. The objective is to determine the key layer whose hidden states are most sensitive to input variations and whose correction most effectively restores the output behavior.

An autoregressive transformer-based language model $G$: $P \rightarrow Y$ typically consists of an embedding layer $E$ followed by a stack of $n$ transformer layers \{$L_1$, $L_2$, \ldots, $L_n$\}. Each transformer layer $L_\ell$ comprises a multi-head self-attention mechanism and a multilayer perceptron (MLP). Given an input prompt $P$, the model first applies the embedding layer $E$ to produce an initial hidden state $h_0$. This representation is then iteratively updated by each layer $L_\ell$, where both the attention heads and the MLP contribute to the transformation of the hidden state:

\vspace{-0.2cm}
\begin{equation}
\label{equ:llmlayer}
h_{\ell} = h_{\ell-1} + \text{MLP}_{\ell} \left( h_{\ell-1} + \text{Att}_{\ell}(h_{\ell-1}) \right)
\end{equation}

$h_{\ell}$ is the hidden state obtained after passing through the $L_\ell$.

Given a pair of prompts (an original prompt $P_{ori}$ and a perturbed prompt $P_{pert}$), we follow three steps to find the key layer responsible for the result variation under the perturbation:

\noindent\ding{182} \textbf{Setup.} We first compute the baseline performance of $P_{ori}$ and $P_{pert}$ using the pass@k metric~\cite{chen2021evaluating} by generating outputs from the model G and checking functional correctness (e.g., via unit tests). Let $\text{Acc}_{ori}$ and $\text{Acc}_{pert}$ denote the pass@1 for $P_{ori}$ and $P_{pert}$.

\noindent\ding{183} \textbf{Layer-wise Intervention.} To ensure a fair comparison and stable intervention, we construct a mini-batch that includes one instance of $P_{ori}$ and multiple instances (e.g., five copies) of $P_{pert}$. \dx{This design allows us to stably compute restoration improvement (as defined in Equation~\ref{equ:restorationImprovement}) across all layers by averaging over repeated forward passes.} The batch is tokenized using left padding, ensuring alignment of token positions across all samples. We input the batch into the model $G$.
For each transformer layer $L_\ell$, where $\ell\in \{1, 2, \ldots, N\}$, we intervene in the forward pass as follows:
\begin{itemize}[leftmargin=0.5cm]

\item During the forward pass, we modify the hidden states of $P_{pert}$ at layer $L_\ell$ by replacing them with the corresponding hidden states from $P_{ori}$:

\vspace{-0.2cm}
\begin{equation}
h_{\ell}^{(i)}(t) \leftarrow h_{\ell}^{ori}(t), \quad \forall i \in \{1, \ldots, B\}, \, \forall t \in \mathcal{T}
\end{equation}

$h_{\ell}^{(i)}(t)$ denotes the hidden state of the i-th $P_{pert}$ at layer $L_\ell$ and token position t; $h_{\ell}^{ori}(t)$ is the corresponding hidden state of $P_{ori}$. $\mathcal{T}$ represents the set of all non-padding token positions, and $B$ is the number of perturbed samples in the batch.
\item The modified hidden states are then propagated forward through layers $L_{\ell+1}$ to $L_N$ to generate output sequences based on the intervened representation.
\item We then run the model from layer $L_{\ell+1}$ onward to generate outputs for the perturbed inputs with patched hidden states. For each generated output, we decode the predicted code, normalize its format, and evaluate it using functional test cases. Let $\text{Acc}_{\ell}^{\text{patched}}$ denote the pass@1 accuracy under this patched configuration at layer $L_\ell$.

\end{itemize}

\noindent\ding{184} \textbf{Key Layer Selection.} To quantify the effectiveness of each intervention, we define the restoration improvement at layer $L_\ell$ as:

\vspace{-0.2cm}
\begin{equation}
\label{equ:restorationImprovement}
\text{Restoration Improvement}_\ell = \frac{\text{Acc}_{\ell}^{\text{patched}} - \text{Acc}_{pert}}{\text{Acc}_{ori} - \text{Acc}_{pert}}
\end{equation}

This ratio captures how much of the accuracy gap between $P_{ori}$ and $P_{pert}$ is recovered by intervention at layer $L_\ell$, normalized to the maximum possible improvement.
We define the key layer $L_{\ell^*}$ as the one with the highest restoration improvement.
If multiple layers achieve the highest restoration improvement, we adopt the approach proposed by Wang et al.~\cite{wang2024detoxifying}.\dx{ We compute the Euclidean distance between the hidden states of each candidate layer when processing $P_{ori}$ and $P_{pert}$, and select the layer with the greatest distance.}

\vspace{-0.1cm}
\subsection{Layer Editing}

Following the localization of the most critical layer for the degradation of robustness, we proceed to perform parameter-level editing to enhance the robustness of the model to a specific perturbation type. 
Each layer's MLP block in Equation~\ref{equ:llmlayer} is a two-layer neural network~\cite{dai2021knowledge,meng2022locating} and the second neural network can be expressed as:

\vspace{-0.2cm}
\begin{equation}
\text{MLP}_\ell(h_{\ell-1}) = h_\ell^{down} W_\ell^V
\end{equation}

$h_\ell^{down}$ denotes the intermediate activation obtained by applying the first linear transformation and nonlinear activation (e.g., GELU) to the input $h_{\ell-1}$, while $W_\ell^V$ represents the output projection matrix that maps the high-dimensional feedforward features back to the hidden size of the model. Previous work~\cite{dai2021knowledge,meng2022locating,wang2024detoxifying,geva2022transformer,lee2024mechanistic} has demonstrated that $W_\ell^V$ plays a critical role in knowledge routing and is often the most effective target for localized interventions. Consequently, we edit $W_{\ell^*}^V$ in the key layer $L_{\ell^*}$ to improve the robustness of the model $G$.

We adopt a gradient-based parameter update strategy to edit model $G$ over $T$ steps, aligning the hidden states of the perturbed prompt with those of the original prompt at the identified key layer $L_{\ell^*}$, while preserving the model’s behavior on clean inputs. During the editing process, only the parameters of $W_{\ell^*}^V$ are updated, while all other parameters of $G$ remain frozen.

Specifically, given an original prompt $P_{ori}$ and its perturbed counterpart $P_{pert}$, we first pass $P_{ori}$ through the model $G$ and extract the hidden state $h^{ori}_{\ell^*}$ at the identified key layer $L_{\ell^*}$. We then input both $P_{ori}$ and $P_{pert}$ into the model to obtain their respective hidden states at $L_{\ell^*}$, denoted as $h^{ori-new}_{\ell^*}$ and $h^{pert}_{\ell^*}$. Here, $h^{ori-new}_{\ell^*}$ represents the hidden state of $P_{ori}$ recomputed after the parameter update and is used to evaluate the preservation of the original behavior of the model.

To measure the squared distance between the two hidden representations, we utilize the Mean Squared Error (MSE) loss. This loss function is widely used and provides a stable method for aligning continuous vector representations. Given two hidden states $h^{(1)}, h^{(2)} \in \mathbb{R}^{n \times d}$, where $n$ is the number of tokens and $d$ is the hidden dimension, the MSE is defined as:

\vspace{-0.1cm}
\begin{equation}
\text{MSE}(h^{(1)}, h^{(2)}) = \frac{1}{n} \sum_{i=1}^{n} \left\| h_i^{(1)} - h_i^{(2)} \right\|^2
\end{equation}

Using this formulation, we define the total loss for the editing procedure as:

\vspace{-0.1cm}
\begin{equation}
\label{equ:loss}
\mathcal{L}_{total} = \underbrace{\text{MSE}(h^{pert}_{\ell^*}, h^{ori}_{\ell^*})}_{\text{alignment loss}} + \lambda \cdot \underbrace{\text{MSE}(h^{ori-new}_{\ell^*}, h^{ori}_{\ell^*})}_{\text{preservation loss}}
\end{equation}

The first term enforces representational alignment between the perturbed and original prompts. In contrast, the second term regulates the parameter update by penalizing deviations from the original hidden states of $P_{ori}$. The hyperparameter $\lambda$ governs the balance between improving robustness and maintaining the behavior of the original model. Subsequently, we used $\mathcal{L}_{total}$ to edit $W_{\ell^*}^V$ through back propagation:

\vspace{-0.1cm}
\begin{align}
\mathcal{W}^{t+1} &= \left[ W_1^{t+1}, \ldots, W_{\ell^*}^{t+1}, \ldots, W_N^{t+1} \right] \notag \\
&= \left[ W_1^{t}, \ldots, W_{\ell^*}^{t} - \nabla_{W_{\ell^*}^V} \mathcal{L}_{total}, \ldots, W_N^{t} \right],
\end{align}

$\left[ W_1^t, \ldots, W_{\ell^*}^t, \ldots, W_N^t \right] $ are parameters of all layers for G at t-th timestep. $W_{\ell^*}^{t}$ is the parameters within the key layer $L_{\ell^*}$, and $\nabla_{W_{\ell^*}^V} \mathcal{L}_{total}$ is the gradient for $W_{\ell^*}^{t}$ at $t$-th timestep. 
After $T$ gradient update steps, we obtain the final edited parameter set $\mathcal{W}^\prime$, where only $W_{\ell^*}^V$ has been modified.

%% file: catalogue/experimentsetup.tex
\begin{table*}[htbp]
\centering
\captionsetup{skip=4pt}
\caption{Categories and Examples of Perturbations}
\label{tab:categories}
\resizebox{\linewidth}{!} {
\small
\begin{tabular}{llp{11cm}}
\toprule
\textbf{ID} & \textbf{Name} & \textbf{Example (Original $\rightarrow$ Perturbed)} \\
\hline
\multicolumn{3}{l}{\textbf{Addition (A1 - A3)}} \\
\hline
A1 & Extra Space inside Words & Write a python function to \examplebef{check} ... $\rightarrow$ Write a python function to \exampleaft{ch eck} ... \\
A2 & Repeated Words & Write a \examplebef{python} function to check ... $\rightarrow$ Write a \exampleaft{python python} function to check ... \\
A3 & Repeated Chars & Write a \examplebef{python} function to check ... $\rightarrow$ Write a \exampleaft{pyython} function to check ... \\

\hline
\multicolumn{3}{l}{\textbf{Deletion (D1 - D4)}} \\
\hline
D1 & Char Deletion & Write a \examplebef{python} function to check ... $\rightarrow$ Write a \exampleaft{pthon} function to check ... \\
D2 & Preposition Deletion & Write a python function \examplebef{to} check ... $\rightarrow$ Write a python function\exampleaft{\_}check ... \\
D3 & Determiner Deletion & ... to check if \examplebef{the} given number is ... $\rightarrow$ ... to check if\exampleaft{\_}given number is ... \\
D4 & Space Deletion & Write a python \examplebef{function to} check ... $\rightarrow$ Write a python \exampleaft{functionto} check ... \\

\hline
\multicolumn{3}{l}{\textbf{Editing (E1 - E6)}} \\
\hline
E1 & Keyboard Typo & Write a python function to \examplebef{check} ... $\rightarrow$ Write a python function to \exampleaft{chack} ... \\
E2 & Extra Capital Letter & Write a python function to \examplebef{check} ... $\rightarrow$ Write a python function to \exampleaft{cHeck} ... \\
E3 & Grammatical Person Variation & \examplebef{Write} a python function to check ... $\rightarrow$ \exampleaft{Writes} a python function to check ... \\
E4 & Active/Passive Voice Variation & \examplebef{Write} a python function to check ... $\rightarrow$ A python function \exampleaft{is written} to check ... \\
E5 & Word Class Variation & \examplebef{Write} a python function to check ... $\rightarrow$ \exampleaft{Writer} a python function to check ... \\
E6 & Synonym Substitution & Write a python function to \examplebef{check} ... $\rightarrow$ Write a python function to \exampleaft{determine} ... \\

\hline
\multicolumn{3}{l}{\textbf{Swap (S1 - S2)}} \\
\hline
S1 & Swap Adjacent Chars & Write a python \examplebef{function} to check ... $\rightarrow$ Write a python \exampleaft{fucntion} to check ... \\
S2 & Swap Adjacent Words & Write a python \examplebef{function to} check ... $\rightarrow$ Write a python \exampleaft{to function} check ... \\

\hline
\multicolumn{3}{l}{\textbf{Paraphrasing (P1 - P2)}} \\
\hline
P1 & Rephrasing Sentence & Print even numbers from a list of numbers.$\rightarrow$ Given a list of numbers, print the even numbers. \\
P2 & Declarative to Interrogative & Print even numbers from a list of numbers. $\rightarrow$ Can you print even numbers from a list of numbers? \\

\hline
\multicolumn{3}{l}{\textbf{Combinations (C1 - C3)}} \\
\hline
C1 & A1 + E1 & Write a \examplebef{python function} to check ... $\rightarrow$ Write a \exampleaft{py thon fanction} to check ... \\
C2 & A3 + E1 & Write a \examplebef{python} function to \examplebef{check} ... $\rightarrow$ Write a \exampleaft{pytthon} function to \exampleaft{chuck} ... \\
C3 & D1 + E1 & Write a \examplebef{python} function to \examplebef{check} ... $\rightarrow$ Write a \exampleaft{pythn} function to \exampleaft{chevk} ... \\

\bottomrule
\end{tabular}}
\vspace{-1.5em}
\end{table*}

\subsection{Baseline}
To evaluate the effectiveness of our approach, we compare it against four baselines: two robustness enhancement methods and two knowledge editing methods.

\subsubsection{Self-Denoising~\cite{agrawal2025enhancing}} is a prompt-based technique designed to improve the resilience of LLMs to instruction-level perturbations. The method guides the model itself to recover the clean version of a perturbed input before execution. In our experiments, we employ the iterative variant (SDi), where the model is employed multiple times to progressively reduce noise in the input prompt. We select SDi because it outperforms the standard self-denoising approach~\cite{agrawal2025enhancing}. However, it does not improve the inherent robustness of the model but instead relies on optimizing the input prompt. In real-world scenarios, we cannot assume that users will modify or refine their prompts prior to interacting with an LLM.

\subsubsection{LoRA Fine-tuning~\cite{agrawal2025enhancing}} is a robustness enhancement strategy based on representation alignment. It inserts LoRA modules into a frozen LLM and trains them to align the hidden representations of perturbed prompts with those of their unperturbed counterparts. Given pairs of original and perturbed prompts, the model minimizes the cosine distance between their mean-pooled middle-layer representations. The LoRA modules are activated only during perturbed inputs, enabling efficient robustness tuning without full model fine-tuning or ground-truth code supervision.
\subsubsection{DINM~\cite{wang2024detoxifying}} is a knowledge editing method developed for detoxification tasks. It identifies the most toxic layer by comparing the hidden states of safe and unsafe completions for the same adversarial input, and fine-tunes this region using a single safe reference. To preserve general capability, it imposes a constraint on unrelated prompts. While both DINM and our method apply localized parameter updates, DINM depends on gold-standard responses, making it well-suited for tasks with clearly defined output references. In contrast, code generation lacks fixed targets, necessitating improvements in robustness without access to predefined ground-truth outputs.
\subsubsection{ROME~\cite{meng2022locating}} is a knowledge editing method that performs immediate and targeted parameter updates to factual associations in LLMs. It identifies the most influential layer and neuron for a given fact using causal tracing, and applies a rank-one update to the MLP output weights at that layer to inject new knowledge. In our work, we adapt ROME to the robustness enhancement setting by using a perturbed prompt and the corresponding standard answer to guide the edit.

\subsection{Dataset}

To evaluate the effectiveness of our method, we use code generation datasets along with their corresponding perturbed versions. Specifically, we use two widely adopted datasets:

\begin{itemize}[leftmargin=*]

\item \textbf{HumanEval}~\cite{chen2021evaluating} is a benchmark consisting of 164 hand-written Python programming problems. Each problem includes an input prompt, a reference solution, and a set of test cases. The prompt comprises a function signature, a functional description, and several example outputs.

\item \textbf{MBPP}~\cite{austin2021program} consists of 974 Python coding problems, including a manually verified subset of 427 problems curated by the authors. Each problem includes a prompt containing a functional description, a reference solution, and test cases. We leverage the manually verified subset to conduct our study.

\end{itemize}

Based on a literature review and an online survey of practitioners, Chen et al.'s work~\cite{chen2024nlperturbator} categorizes common prompt perturbations and propose an automated framework, \textbf{NLPerturbator}, which applies each type of perturbation to a given set of prompts.
We follow their work and get the perturbed dataset (i.e., HumanEval-R and MBPP-R).
As shown in Table~\ref{tab:categories}, we adopt 20 perturbation types that span lexical and syntactic variations. These perturbations are designed to preserve the original semantics while introducing surface-level deviations that challenge the robustness of models.

To ensure a meaningful robustness evaluation, it is essential to identify those prompts that genuinely cause deviations in the output behavior of models. We first compute the pass rate (i.e., functional correctness measured via test case success) of the original prompt set, denoted as $A cc_{ori}$. For each perturbation type, we then compute $Acc_{pert}$ on the corresponding perturbed prompts. The robustness drop for each prompt is defined as:

\vspace{-0.1cm}
\begin{equation}
\Delta_{robustness} = Acc_{ori} - Acc_{pert}
\end{equation}

However, due to the inherent randomness in LLM generation, small fluctuations in pass rate may not reliably indicate a robustness issue. To mitigate this effect and focus on impactful perturbation prompts, we introduce a filtering threshold $\delta = 0.3$ and retain only those samples for which $\Delta_{robustness} \geq \delta$.
This threshold balances sensitivity and coverage: a lower value may admit noise due to random variation, while a higher threshold (e.g., $\delta = 0.5$) substantially reduces the number of usable samples. 
The filtered subset serves as a perturbation-sensitive benchmark for evaluating robustness-enhancing methods.

\subsection{Evaluation Metrics}
To evaluate the robustness of code LLMs under prompt perturbations, we employ two complementary metrics: pass@1 and Generalized Relative Improvement Ratio (G-RIR).

\textbf{Pass@1} is a standard metric for evaluating the functional correctness of generated code~\cite{chen2021evaluating}. It measures the proportion of generated outputs that pass all test cases on the first attempt. In this study, we adopt pass@1 rather than higher-$k$ variants (e.g., pass@5 or pass@10), as it reflects the practical setting in which users expect a correct solution from a single generation without manual selection or reranking. Formally, let $n$ denote the number of generated samples for a given prompt, $c$ the number of correct generations. The unbiased pass@k metric is defined as:


\begin{equation}
\text{Pass@}k = 1 - \frac{{\binom{n - c}{k}}}{{\binom{n}{k}}}, \quad \text{for } k \leq n
\end{equation}

\textbf{G-RIR} is a metric we propose to quantify the generalization ability of a robustness enhancement strategy. Given a set of $N{+}x$ prompts associated with a specific perturbation type, we use $x$ prompts to update the model (e.g., $x=1$ in our method). We then evaluate the resulting robustness improvement on the remaining $N$ prompts, which exhibit the same type of perturbation but are unseen during the model updating. Formally, G-RIR is defined as:

\begin{equation}
\text{G-RIR} = \frac{1}{N} \sum_{i=1}^{N} \frac{A^{\text{after}}_{\text{pert},i} - A^{\text{before}}_{\text{pert},i}}{A^{\text{before}}_{\text{orig},i} - A^{\text{before}}_{\text{pert},i} + \epsilon}
\end{equation}

$A^{\text{before}}_{\text{pert},i}$ and $A^{\text{after}}_{\text{pert},i}$ represent the pass@1 scores of the $i$-th perturbed prompt before and after applying the robustness enhancement method, respectively. $A^{\text{before}}_{\text{orig},i}$ denotes the pass@1 score of the corresponding original prompt. A small constant $\epsilon$ is added to the denominator to prevent division by zero. The numerator captures the absolute performance improvement on the perturbed prompt due to the enhancement method, while the denominator reflects the maximum possible improvement. This metric quantifies the average restoration effectiveness, normalized by the robustness gap prior to editing. A higher G-RIR indicates stronger and more consistent robustness improvements across tasks within a given perturbation type.

\subsection{Experimental Setting}
\label{sec:experimentalSetting}
All experiments are conducted using PyTorch and HuggingFace Transformers on a machine equipped with an NVIDIA A800 GPU (80 GB memory). For evaluation, we use two widely adopted open-source code LLMs: \texttt{CodeLlama-7b} and \texttt{Qwen2.5-Coder-7B}, both loaded in half-precision (i.e., FP16) with left-padded tokenization.

During robustness editing, we set the learning rate to $1 \times 10^{-3}$ and allow up to 20 editing steps. The Adam optimizer is used with a weight decay of $1 \times 10^{-5}$. Early stopping is applied: the editing terminates if the loss does not improve for three consecutive steps. The regularization coefficient $\lambda$ for preserving the original hidden state is set to 0.1. Only the MLP output projection matrix $W_{\ell^*}^V$ at the identified key layer $L_{\ell^*}$ is updated, while all other model parameters remain frozen.

For code generation, we set the sampling temperature to 0.2 to reduce randomness and ensure evaluation stability. We apply our editing method to one perturbed prompt per perturbation type and assess its generalization to other prompts of the same type.

To evaluate robustness generalization, we adopt a leave-one-in evaluation protocol within each perturbation type. Specifically, for a given type (e.g., C3 or D2) in Table~\ref{tab:categories}, each task is used in turn to perform model editing or enhancement, after which the updated model is evaluated on the remaining tasks in the same category. We report performance using functional correctness (Pass@1) and generalization ability (G-RIR) to quantify robustness improvement.

%% file: catalogue/results.tex
\begin{table*}[ht]
\vspace{-0.1em}
\centering
\captionsetup{skip=3pt}
\caption{Comparison of Pass@1 across models, datasets, and perturbation types. Raw scores are followed by percentage improvements over the perturbed baseline. Higher is better.}
\small
\begin{tabular}{l|cc|cc|cc|cc|c}
\hline
 & \multicolumn{2}{c|}{CodeLlama + HumanEval} & \multicolumn{2}{c|}{CodeLlama + MBPP} & \multicolumn{2}{c|}{QWenCoder + HumanEval} & \multicolumn{2}{c|}{QWenCoder + MBPP} &  \\
\multirow{-2}{*}{\textbf{Method}} & C3↑ & D2↑ & C3↑ & D2↑ & C3↑ & D2↑ & C3 & D2↑ &\multirow{-2}{*}{Avg↑}\\
\hline
\footnotesize{Base LLM (non-perturbation)} & 0.600 & 0.583 & 0.688 & 0.710 & 0.739 & 0.625 & 0.780 & 0.769&- \\
\footnotesize{Base LLM (perturbation)} & 0.220 & 0.150 & 0.208 & 0.190 & 0.300 & 0.238 & 0.190 & 0.375&- \\
\hline
Self-Denoising & 0.062 {\small (-72\%)} & 0.113 {\small (-25\%)} & 0.033 {\small (-84\%)} & 0.060 {\small (-68\%)} & 0.014 {\small (-95\%)} & 0.180 {\small (-24\%)} & \textbf{0.444} {\small (\textbf{134\%})} & 0.505 {\small (35\%)}&-25\% \\
LoRa       & 0.294 {\small (34\%)} & 0.283 {\small (89\%)} & 0.329 {\small (58\%)} & 0.327 {\small (72\%)} & 0.379 {\small (26\%)} & 0.198 {\small (-17\%)} & 0.427 {\small (125\%)} & 0.477 {\small (27\%)}&52\% \\
ROME       & 0.196 {\small (-11\%)} & 0.280 {\small (87\%)} & 0.129 {\small (-38\%)} & 0.201 {\small (6\%)} & 0.400 {\small (33\%)} & 0.307 {\small (29\%)} & 0.299 {\small (57\%)} & 0.546 {\small (46\%)}&26\% \\
DINM       & 0.227 {\small (3\%)} & 0.223 {\small (49\%)} & 0.283 {\small (36\%)} & 0.256 {\small (35\%)} & 0.177 {\small (-41\%)} & 0.166 {\small (-30\%)} & 0.392 {\small (106\%)} & 0.408 {\small (9\%)}&21\% \\
\hline
\textbf{\appname (Ours)}  & \textbf{0.304} {\small (\textbf{38\%})} & \textbf{0.287} {\small (\textbf{91\%})} & \textbf{0.340} {\small (\textbf{63\%})} & \textbf{0.331} {\small (\textbf{74\%})} & \textbf{0.422} {\small (\textbf{41\%})} & \textbf{0.354} {\small (\textbf{49\%})} & 0.396 {\small (108\%)} & \textbf{0.548} {\small (\textbf{46\%})}& \textbf{64\%}\\
\hline
\end{tabular}
\label{tab:rq1_passk}
\vspace{-0.4em}
\end{table*}

\begin{table*}[ht]
\centering
\captionsetup{skip=3pt}
\caption{Comparison of G-RIR across models, datasets, and perturbation types. Higher is better.}
\begin{tabular}{l|cc|cc|cc|cc|c}
\hline
 & \multicolumn{2}{c|}{CodeLlama + HumanEval} & \multicolumn{2}{c|}{CodeLlama + MBPP} & \multicolumn{2}{c|}{QWenCoder + HumanEval} & \multicolumn{2}{c|}{QWenCoder + MBPP}& \\
 \multirow{-2}{*}{\textbf{Method}}& C3↑ & D2↑ & C3↑ & D2↑ & C3↑ & D2↑ & C3 & D2↑&\multirow{-2}{*}{Avg↑} \\
\hline
Self-Denoising & - & - & - & - & - & - & - & -&- \\
LoRa       & 0.2407 & 0.3139 & 0.3117 & 0.3776 & 0.1218 & -0.0818 & \textbf{0.4660} & 0.3445&0.2618 \\
ROME           & 0.0874 & 0.2833 & -0.2070 & 0.0063 & 0.2693 & 0.1934 & 0.2375 & 0.4891&0.1699 \\
DINM           & -0.0072 & 0.1500 & 0.1685 & 0.1207 & -0.2829 & -0.2083 & 0.3448 & 0.1236&0.0510 \\
\hline
\textbf{\appname (Ours)}  & \textbf{0.2956} & \textbf{0.3333} & \textbf{0.3414} & \textbf{0.3786} & \textbf{0.3502} & \textbf{0.3051} & 0.4332 & \textbf{0.5120}&\textbf{0.3687} \\
\hline
\end{tabular}
\label{tab:rq1_grir}
\vspace{-0.5em}
\end{table*}
In this paper, we aim to answer the following four research questions:

\noindent\textbf{RQ.1 (Effectiveness)} How effective is our method compared to existing robustness enhancement and knowledge editing approaches?

\noindent\textbf{RQ.2 (Generality)} How does the effectiveness of our method vary across different types of prompt perturbations?

\noindent\textbf{RQ.3 (Ablation Study)} How do the analyses described in Section~\ref{sec:methodology} contribute to the overall effectiveness?

\noindent\textbf{RQ.4 (Layer Distribution Analysis)} How are robustness-sensitive layers distributed across models and perturbation types?

\vspace{-0.1cm}
\subsection{RQ1: Effectiveness}

To evaluate the effectiveness of \appname, we compare our method with several baselines, including two robustness enhancement techniques (i.e., Self-Denoising and LoRA fine-tuning) and two knowledge editing methods (i.e., ROME and DINM). \dx{We assess performance on two representative perturbation types, C3 and D2, which represent different levels of perturbation intensity: D2 introduces a single localized modification, whereas C3 involves multiple dispersed changes.}
We report both absolute functional accuracy (Pass@1) and robustness generalization (G-RIR) under the leave-one-in evaluation protocol (see Section~\ref{sec:experimentalSetting}).

\noindent\faThumbsUp ~\textbf{Pass@1 Comparison}: 
\appname exhibits strong potential in enhancing model robustness. As shown in Table~\ref{tab:rq1_passk}, it achieves the highest Pass@1 scores in 7 out of 8 settings. In all cases, \appname outperforms the unedited model on perturbed prompts, improving the ability of models to generate functionally correct code despite input perturbations.
Notably, the Self-Denoising approach performs poorly in most settings, except for \texttt{QWenCoder+MBPP}, contrasting with its previously reported effectiveness~\cite{agrawal2025enhancing}.
This discrepancy may be attributed to two factors: \ding{182} unlike like \texttt{QwenCoder}, \texttt{CodeLlama} is not instruction-tuned, which limits its ability to revise perturbed prompts based on meta-instructions; \ding{183} prompts in HumanEval average 231.6 characters in length, significantly longer than those in MBPP (88.2 characters), which increases the likelihood of semantic drift during denoising.

\noindent\faThumbsUp ~\textbf{G-RIR Comparison}: \appname exhibits strong generalization capabilities for robustness enhancement. As shown in Table~\ref{tab:rq1_grir}, it achieves the highest G-RIR scores in nearly all evaluated settings. For instance, \appname achieves a G-RIR of 0.5120 under the \texttt{Qwen + MBPP + D2} configuration, significantly outperforming all baselines.
Self-Denoising is excluded from the G-RIR comparison as it is designed to enhance robustness on a single instance and does not support generalization across tasks within the same perturbation category.
Additionally, some methods (e.g., ROME and DINM) yield negative G-RIR values in certain settings, suggesting that editing with a single standard answer may lead to overfitting and reduced robustness on the same perturbation.

\begin{center}
    \resizebox{\linewidth}{!}{
\begin{tabular}{l!{\vrule width 1pt}p{0.9\columnwidth}}
    \makecell{{\LARGE \faLightbulbO}}  &\textbf{Answer to RQ1:}
    \appname achieves the best overall performance, with a 63\% average improvement in Pass@1 over the base LLM on perturbed prompts and a 23\% relative gain over the best baseline (LoRA). It also obtains the highest average G-RIR score (0.37), representing a 41\% improvement over the best-performing baseline. This suggests that robustness improvements from a single edit generalize well to other tasks within the same perturbation category.\\
\end{tabular}}
\end{center}

\vspace{-0.1cm}
\subsection{RQ2: Generality}
\begin{table*}[ht]
\captionsetup{skip=3pt}
\centering
\caption{G-RIR scores (\%) across different types of prompt perturbations. Higher is better.}
\label{tab:rq2_g_rir}
\resizebox{\textwidth}{!}{
\begin{tabular}{l|cccccccccccccccccccc|c}
\hline
\textbf{Perturbation Type} & A1 & A2 & A3 & D1 & D2 & D3 & D4 & E1 & E2 & E3 & E4 & E5 & E6 & S1 & S2 & P1 & P2 & C1 & C2 & C3&Avg \\
\hline
CodeLlama+humaneval & 22 & 17 & 11 & 37 & 33 & 50 & 18 & 30 & 29 & 11 & 3 & 23 & 39 & 23 & 6 & 7 & 19 & 46 & 16 & 30&24  \\
CodeLlama+MBPP       & 20 & 72 & 40 & 14 & 38 & 34 & 52 & 45 & 50 & 40 & 35 & 53 & 44 & 56 & 24 & 27 & 35 & 19 & 37 & 34&38  \\
Qwen+humaneval       & 30 & 14 & 34 & 31 & 31 & 17 & 17 & 16 & 42 & 50 & 34 & 22 & 23 & 13 & 26 & 10 & 9 & 11 & 9 & 35&24  \\
Qwen+MBPP            & 43 & 67 & 52 & 17 & 51 & 29 & 49 & 38 & 22 & 35 & 21 & 39 & 27 & 29 & 38 & 19 & 34 & 10 & 36 & 43&35 \\
\hline
Average            & 29 & 43 & 34 & 25 & 38 & 33 & 34 & 33 & 36 &34 & 23 & 34 & 33 & 30 & 24 & 16 & 24 & 22 & 25 & 36&30 \\
\hline
\end{tabular}
}
\vspace{-0.5em}
\end{table*}

\begin{table*}[ht]
\captionsetup{skip=3pt}
\centering
\caption{Cross-Category Generalization: G-RIR (\%) on Unseen Perturbation Types After Editing on C3 (HumanEval+QwenCoder)}
\label{tab:cross_generalization}
\resizebox{\textwidth}{!}{
\begin{tabular}{l|ccccccccccccccccccc|c}
\hline
\textbf{Perturbation Type} & A1 & A2 & A3 & D1 & D2 & D3 & D4 & E1 & E2 & E3 & E4 & E5 & E6 & S1 & S2 & P1 & P2 & C1 & C2 &Avg \\
\hline
\textbf{G-RIR} & 24 & 19 & 28 & 23 & 21 & 2 & 16 & 9 & 32 & 38 & 25 & 19 & 20 & 3 & 19 & 10 & 15 & 11 & 7 & 23  \\

\hline
\end{tabular}
}
\vspace{-1em}
\end{table*}

To evaluate the generalization capability of our method across different categories of prompt perturbations, we calculate the G-RIR for each perturbation type listed in Table~\ref{tab:categories}.
Table~\ref{tab:rq2_g_rir} reports the G-RIR scores for 20 perturbation types, including additions (A1–A3), deletions (D1–D4), edits (E1–E6), swaps (S1–S2), paraphrases (P1–P2), and co-occurring categories (C1–C3). Our key findings are summarized as follows:

\noindent\faLightbulbO~\textbf{Finding 1.} Our method achieves an average G-RIR score of 30\% and consistently yields positive G-RIR scores across all perturbation types, indicating strong generalization.

\noindent\faLightbulbO~\textbf{Finding 2.}
\appname demonstrates stronger generalization on word-level perturbations compared to sentence-level rephrasings. Specifically, perturbation types in the Addition (A1–A3), Deletion (D1–D4), and Editing (E1–E6) categories generally yield higher G-RIR scores, with many cases exceeding 30\%. These perturbations introduce changes that preserve the original semantic intent. In contrast, rephrasing (P1, P2) and swapping (S1, S2) perturbations are more challenging. For example, P1 yields the lowest average G-RIR scores of 16\%. These types often involve sentence-level rewording or syntactic restructuring, which may cause the model to shift its internal attention or misinterpret the functional objective of the prompt.

\noindent\faLightbulbO~\textbf{Finding 3.}
The effectiveness of our method varies across datasets. The average G-RIR score on MBPP is consistently higher than that on HumanEval for both \texttt{CodeLlama} and \texttt{Qwen}. For example, \texttt{CodeLlama} achieves an average G-RIR of 38\% on MBPP compared to 24\% on HumanEval, while \texttt{Qwen} scores 35\% on MBPP and 24\% on HumanEval. This discrepancy may be attributed to the nature of the tasks: MBPP prompts are generally shorter and more templated, whereas HumanEval problems tend to be more descriptive and structurally diverse. Consequently, \appname demonstrates better generalization on simpler, more regular prompts.

\noindent\dx{\faLightbulbO~\textbf{Finding 4.}
To evaluate whether model editing tailored to a specific perturbation category can generalize to others, we conduct an additional cross-category evaluation. Specifically, the model is edited using only C3 type examples under the \texttt{HumanEval+QwenCoder} setting and then tests across all other perturbation types. As shown in Table~\ref{tab:cross_generalization}, the model achieves an average G-RIR improvement of 18\% on unseen perturbation types, compared to 24\% when both editing and evaluation are performed within the same category (as reported in Table~\ref{tab:rq2_g_rir}). These results suggest that \appname does not overfit to a specific perturbation category and can generalize to a broader range of prompt variations.}

\begin{center}
    \resizebox{\linewidth}{!}{
\begin{tabular}{l!{\vrule width 1pt}p{0.9\columnwidth}}
    \makecell{{\LARGE \faLightbulbO}}  &\textbf{Answer to RQ2:}
    Our method demonstrates strong generalization across various categories of perturbations, achieving an average improvement of 30\%. It performs better on simpler perturbations (e.g., word-level modifications) than on more complex sentence-level rephrasings. \dx{Moreover, \appname preserves robustness improvements even when trained on a single perturbation type and evaluated across others, achieving an average cross-type G-RIR of 18\%.}\\
\end{tabular}}
\end{center}

\begin{table*}[ht]
\centering
\captionsetup{skip=3pt}
\caption{G-RIR scores for the ablation variants of our method under different settings. Higher is better.}
\label{tab:ablation}
\begin{tabular}{l|cc|cc|cc|cc}
\hline
 & \multicolumn{2}{c|}{CodeLlama + HumanEval} & \multicolumn{2}{c|}{CodeLlama + MBPP} & \multicolumn{2}{c|}{QWenCoder + HumanEval} & \multicolumn{2}{c}{QWenCoder + MBPP} \\
 \multirow{-2}{*}{\textbf{Method}}& C3 & D2 & C3 & D2 & C3 & D2 & C3 & D2 \\
\hline
w/o Layer Localization & 0.2296$\downarrow$  & 0.2306  & 0.3003$\downarrow$  & 0.2364$\downarrow$ &0.3249$\downarrow$ & 0.2594$\downarrow$  & 0.3074$\downarrow$  & 0.4838$\downarrow$ \\
w/o Early Stopping       & 0.2593  & 0.2278$\downarrow$  & 0.3305 & 0.2824  & 0.3540 & 0.2634  & 0.3398  & 0.4969  \\
w/o Preserve Loss ($\mathcal{L}_{\text{preserve}}$)   & 0.2481  & 0.2528  & 0.3469 & 0.3042  & 0.3383 &0.2693  & 0.3467  & 0.4870  \\
\hline
\textbf{CREME\_Full (Ours)}  & 0.2768 & 0.3194 & 0.3414 & 0.3786 & 0.3502 & 0.3275 & 0.4766 & 0.5120 \\
\hline
\end{tabular}
\end{table*}
\vspace{-0.1cm}
\subsection{RQ3: Ablation Study}
To investigate the contribution of each component to the effectiveness of \appname, we create the following variants:
\begin{itemize}[leftmargin=*]
    \item \textbf{CREME\_Full}: The complete version of our method, exactly as Section~\ref{sec:methodology} illustrated.
    \item \textbf{w/o Layer Localization}: This variant skips the causal tracing step and instead randomly selects a middle layer for editing. This helps assess the importance of precisely locating the key layer responsible for robustness degradation.
    \item \textbf{w/o Early Stopping}: Early stopping is disabled, allowing all editing steps to be executed regardless of the loss trend. This removes the safeguard designed to prevent overfitting.
    \item \textbf{w/o Preserve Loss}: This variant removes the preservation loss term, which constrains the model to maintain its behavior on clean inputs, thereby allowing the model to deviate from its original performance.
\end{itemize}
Table~\ref{tab:ablation} reports the G-RIR scores of \appname and its ablated variants across four evaluation settings. Overall, CREME\_Full consistently achieves the highest G-RIR scores, validating the contribution of each individual component. Among the ablation variants, removing the layer localization module results in the greatest performance degradation (20.4\% on average), underscoring the critical importance of locating the key layer related to robustness.
Disabling early stopping results in a 14.2\% average decline, emphasizing its importance in preventing overfitting during model editing. Removing the preservation loss term causes a 12.8\% reduction, indicating that preserving model behavior on clean inputs is also crucial for achieving robust generalization.

\begin{center}
    \resizebox{\linewidth}{!}{
\begin{tabular}{l!{\vrule width 1pt}p{0.9\columnwidth}}
    \makecell{{\LARGE \faLightbulbO}}  &\textbf{Answer to RQ3:}
    All components of \appname are essential to its effectiveness, with layer localization contributing most significantly (–20.4\% G-RIR).\\
\end{tabular}}
\end{center}

\vspace{-0.1cm}
\subsection{RQ4: Layer Distribution Analysis}
\label{sec:layerDistribution}
\begin{figure*}[h]
\captionsetup{skip=2pt}
  \centering  \includegraphics[width=\linewidth]{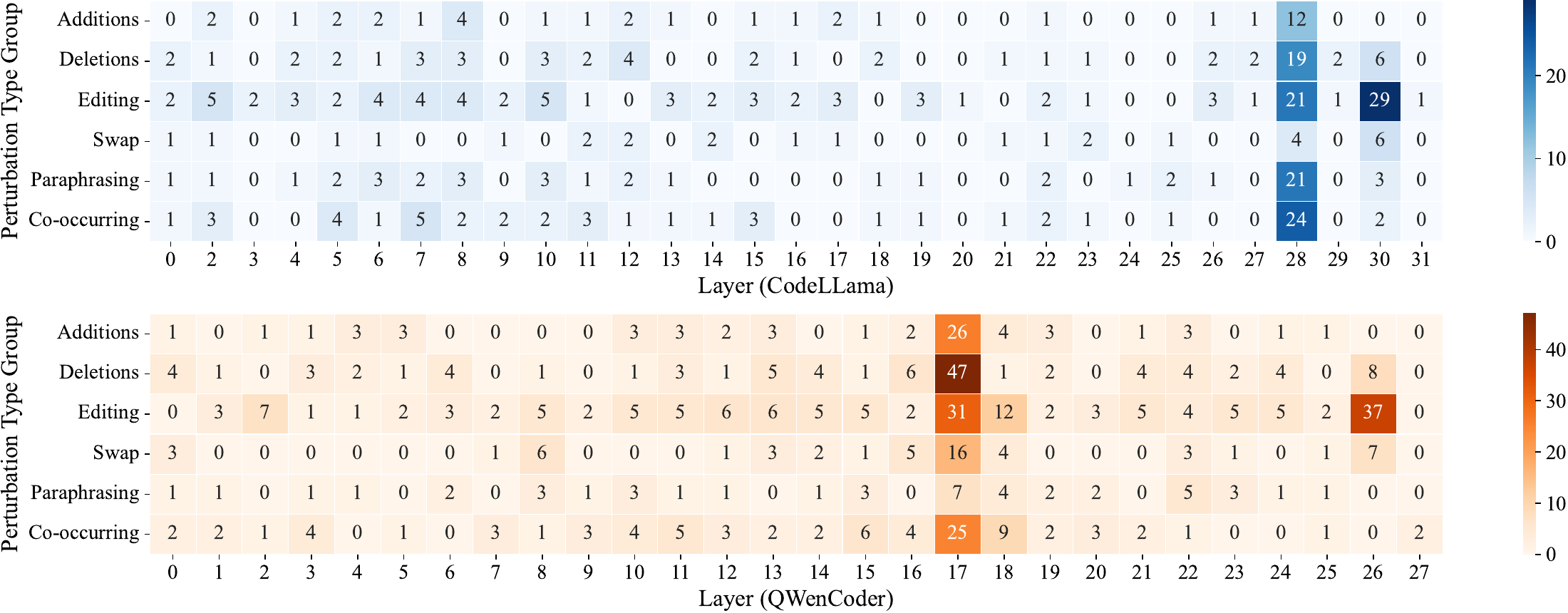}
  \caption{Key Layer Distribution by Perturbation Type Group}
    \label{fig:layerDistribution}
\vspace{-1em}
\end{figure*}
To understand how robustness-related information is localized across the model’s architecture, we analyze the distribution of key layers selected by our causal tracing procedure under different perturbation types and models. Figure~\ref{fig:layerDistribution} presents the heatmaps of key layer frequencies for the \texttt{CodeLlama-7b} and \texttt{Qwen2.5-Coder-7B} models, grouped by six perturbation categories.
\dx{The ``key layer frequency'' refers to how many times each transformer layer is selected as the most robustness-sensitive layer across all editing cases (2 datasets × 20 perturbation types).}

\noindent\faHandORight\ \textbf{Model Dimension:} Robustness-sensitive layers are concentrated in specific regions of the network and vary significantly across models. In the case of \texttt{CodeLlama}, key layers are generally located in the later stages of the model and exhibit a more dispersed distribution (mean = 18.93, standard deviation = 10.08). Notably, layers 28 and 30 are the most frequently identified, together accounting for over 40\% of all key layers. In contrast, \texttt{QWenCoder} exhibits a more centralized distribution (mean = 15.76, standard deviation = 6.57), with a strong concentration at layer 17, suggesting a more localized region for robustness-relevant computations. Despite differences in the number of layers and architectural design, both models consistently exhibit a concentration of robustness-sensitive layers in the middle-to-deep regions of their networks. These findings support the hypothesis that the robustness-related region is not uniformly distributed across the model but is instead governed by specific layers that encode perturbation-sensitive representations.  Identifying and targeting these regions is therefore essential for effective robustness enhancement.

\noindent\faHandORight\ \textbf{Perturbation Dimension:}
Most perturbation types exhibit similar distributions of robustness-sensitive key layers, while certain categories display distinct patterns. In the case of \texttt{CodeLlama}, most perturbation types (e.g., additions, deletions, and paraphrasing) tend to concentrate key layers around Layer 28. However, the Editing category exhibits a distinct pattern by shifting the key layer distribution toward Layer 30. This shift suggests that editing perturbations may engage robustness-sensitive mechanisms in deeper layers, potentially due to their grammatical and lexical complexity that demands more semantic-level processing.
A similar trend is observed in \texttt{QWenCoder}, where most perturbation types concentrate around Layer 17. However, Editing perturbations notably shift the distribution toward Layer 26.
These results suggest that while key layers remain relatively consistent within similar perturbation types, distinct perturbation categories can activate robustness-sensitive regions in different parts of the model.

Overall, these findings suggest that robustness-sensitive regions are not uniformly distributed across all layers but tend to cluster within specific parts of the network. The locations of these regions vary depending on both the model architecture and the type of perturbation. This non-uniformity implies the existence of specialized subregions within the model that are particularly responsible for processing robustness-related information. 
These observations empirically validate our layer localization strategy and offer valuable insights for designing more effective robustness enhancement techniques.

\begin{center}
    \resizebox{\linewidth}{!}{
\begin{tabular}{l!{\vrule width 1pt}p{0.9\columnwidth}}
    \makecell{{\LARGE \faLightbulbO}}  &\textbf{Answer to RQ4:}
    Robustness-sensitive layers are unevenly distributed and tend to cluster within specific layers of the model, with their locations varying across model architectures and perturbation types.\\
\end{tabular}}
\end{center}

%% file: catalogue/discussion.tex
\begin{table*}[t]
\centering
\captionsetup{skip=2pt}
\caption{Code accuracy (\textit{pass@k}) on original inputs before and after editing. $\Delta$ denotes the change in accuracy after editing.}
\captionsetup{skip=3pt}
\label{tab:editing_drop}
\begin{tabular}{lccccccc}
\toprule
\textbf{Model} & \textbf{Dataset} & \textbf{pass@1 (Orig)} & \textbf{pass@1 (Edited)} & $\Delta$ & \textbf{pass@10 (Orig)} & \textbf{pass@10 (Edited)} & $\Delta$ \\
\midrule
CodeLlama-7b & HumanEval & 30.43\% & 30.25\% & -0.17\% & 43.90\% & 44.62\% & 0.72\% \\
CodeLlama-7b & MBPP      & 51.69\% & 51.67\% & -0.02\% & 67.33\% & 67.35\% & 0.02\% \\
Qwen2.5-Coder-7B   & HumanEval & 45.67\% & 46.35\% & 0.68\% & 62.80\% & 62.56\% & -0.24\% \\
Qwen2.5-Coder-7B   & MBPP      & 63.81\% & 63.59\% & -0.22\% & 79.10\% & 79.58\% & 0.48\% \\
\bottomrule
\end{tabular}
\vspace{-1em}
\end{table*}

\subsection{Impact of Editing on Model Performance}
While \appname effectively enhances robustness against perturbed prompts, it is crucial to evaluate whether the editing process inadvertently degrades model performance on clean data. 
We conduct a case study by randomly sampling 40 editing tasks from the dataset used in each experimental configuration. To ensure balanced coverage across perturbation types, the sampled tasks include two instances for each perturbation type listed in Table~\ref{tab:categories}.
For each task, we apply the \appname editing procedure and then evaluate the edited model on the corresponding original dataset (i.e., HumanEval or MBPP) with clean prompts.
Table~\ref{tab:editing_drop} presents the comparison of pass@1 and pass@10 accuracy before and after editing. The results indicate that the performance of \appname remains stable on clean inputs, with most variations within a $\pm$1\% margin. 

Overall, these results demonstrate that \appname achieves robustness enhancement with minimal impact on the original capabilities of the model.

\subsection{Costs of \appname}
\dx{To assess the practical feasibility of CREME, we conducted a cost evaluation on the \texttt{HumanEval+QwenCoder} setting using the C3 perturbation category. Table~\ref{tab:costs} presents a comparison of \appname with four representative baseline methods in terms of localization, training, and testing time.}

\dx{The layer localization step of \appname takes an average of 495s (8.3 minutes). While this step introduces additional overhead, it is crucial for identifying the most robustness-sensitive layer. In contrast, the two model editing methods, ROME and DINM, rely on either fixed layers or hidden-state distance, neither of which is robustness-aware. As demonstrated in Table~\ref{tab:ablation}, our localization strategy substantially improves robustness over fixed-layer editing, justifying the additional cost. Regarding training time, CREME requires only 5.3s, which is significantly less than DINM (15.8s) and comparable to the others. For inference, CREME’s average test time (3.0s) is competitive and markedly faster than Self-Denoising (29.4s), which iteratively adjusts prompts during generation. Given that models are typically deployed across many downstream tasks after editing, minimizing test time is more critical.}

\dx{To reduce the overhead of layer localization, we propose a lightweight variant that limits localization to a small set of frequently identified robustness-sensitive layers from Section~\ref{sec:layerDistribution} (e.g., layers 28/30 in \texttt{CodeLlama} and 17/26 in \texttt{QwenCoder}), which significantly lowers localization time cost while maintaining effectiveness.}
\begin{table}[ht]
\centering
\captionsetup{skip=3pt}
\caption{Comparison of Time Costs (s) for CREME and Baselines on HumanEval+QwenCoder (C3 perturbation)}
\begin{tabular}{l|ccc}
\toprule
 \textbf{Method} & \footnotesize{Localization Time (s)} & \small{Train Time (s)} & \small{Test Time (s)}  \\
\midrule
Self-Denoising & - & - & 29.4  \\
LoRa       & – & 9.2 & \textbf{2.8}  \\
ROME           & – & 7.59 & 3.4  \\
DINM           & 0.2 & 15.8 & 3.9  \\
\midrule
\textbf{\appname}  & 494.5 & \textbf{5.3} & 3.0  \\
\bottomrule
\end{tabular}
\label{tab:costs}
\end{table}

\subsection{Hyperparameter Sensitivity of $\lambda$}

\dx{The hyperparameter $\lambda$ in the Equation~\ref{equ:loss} controls the trade-off between the preservation loss and the alignment loss. To guide practitioners in choosing an appropriate value for $\lambda$, we conducted a sensitivity analysis using the C3 perturbation category under the \texttt{HumanEval+QwenCoder} setting. As shown in Table~\ref{tab:lambda_sensitivity}, both G-RIR and Pass@1 scores improve as $\lambda$ increases from 0.01 to 0.1, peaking at $\lambda = 0.1$. Beyond this point, performance begins to decline. These findings indicate that $\lambda = 0.1$ provides a robust trade-off and can serve as a strong default for future applications of our framework.}

\begin{table}[h]
\centering
\caption{Sensitivity Analysis of $\lambda$ on HumanEval+QwenCoder (C3 perturbation)}
\label{tab:lambda_sensitivity}
\begin{tabular}{l|cccccc}
\toprule
\textbf{$\lambda (lambda)$} & \textbf{0.01} & \textbf{0.05} & \textbf{0.1} & \textbf{0.2} & \textbf{0.5} & \textbf{1.0} \\
\midrule
G-RIR (\%)  & 32.1 & 33.2 & \textbf{35.0} & 30.3 & 32.8 & 27.6 \\
Pass@1 (\%) & 41.4 & 41.9 & \textbf{42.2} & 40.8 & 41.2 & 39.6 \\
\bottomrule
\end{tabular}
\vspace{-1em}
\end{table}
\subsection{Threats to Validity}
\subsubsection{Internal Validity}
A potential threat arises from the accuracy of our key layer localization strategy based on causal tracing. Although we identify robustness-sensitive layers using both restoration improvement metric and L2-based refinement, there remains a risk of misidentification due to randomness in model behavior or dataset noise. To mitigate this, we aggregate results across multiple tasks and perturbation types. 
Another concern is the stability of evaluation. Since code generation involves stochastic sampling (i.e., with temperature = 0.2), the results may exhibit slight variance. We address this by standardizing generation settings and averaging results across multiple completions per prompt.
\subsubsection{External Validity}
Our study is conducted using two open-source code generation models (i.e., \texttt{CodeLlama} and \texttt{QwenCoder}) and two widely used benchmarks (i.e., HumanEval and MBPP). While these choices reflect diversity in both model architectures and prompt styles, the generalizability of our approach to other LLMs (e.g., \texttt{GPT-4} and \texttt{DeepSeek}) or to domains beyond code generation (e.g., natural language question answering or summarization) remains unverified.
Our concern is that commercial LLMs are continuously updated and may become outdated or permanently inaccessible. In contrast, open-source LLMs offer stable access once released, and their historical versions can be revisited. We select code LLMs for our study because they can generate code directly without requiring additional fine-tuning on external code datasets.
Furthermore, the perturbations employed are derived from the NLPerturbator taxonomy, which may not comprehensively represent robustness challenges under more adversarial or out-of-distribution conditions. Extending our evaluation to include broader and more aggressive perturbation strategies will be an important direction for future work.

%% file: catalogue/relatedwork.tex
In this section, we present a comprehensive review of prior research on LLM robustness and the knowledge editing methods. Beyond LLM robustness, SE work has studied software security and practical deployment issues~\cite{liu2025empirical,li2022poison,chen2025diffploitfacilitatingcrossversionexploit,pan2022automated,yang2024federated}.
\subsection{LLM Robustness}
Despite achieving impressive performance in increasingly sophisticated tasks~\cite{liu2023fill,feng2024prompting,geng2024large}, LLMs remain sensitive to input perturbations. While humans are generally robust to minor variations in natural language task descriptions~\cite{walkington2019effect}, LLMs often produce significantly different outputs in response to such changes~\cite{moradi2021evaluating,honarvar2025turbulence}, highlighting their limited robustness.
Wang et al.~\cite{wang2021adversarial} and Zhu et al.~\cite{zhu2023promptrobust} introduce benchmark suites such as Adversarial GLUE and PromptRobust, which systematically evaluate LLMs under adversarial or subtly perturbed prompts.

In the domain of code generation, a series of studies have revealed that current code LLMs are sensitive to even minor variations~\cite{wang2022recode,shirafuji2023exploring,lam2025codecrash}. Chen et al.~\cite{chen2024nlperturbator} categorize 18 types of natural language perturbations along with three co-occurring combinations and develop NLPerturbator, a framework targeting real-world prompt perturbations. Their findings show that such perturbations can substantially degrade code generation performance(e.g., up to 21.2\%, and 4.8\% to 6.1\% on average). Lin et al.~\cite{lin2025robunfr} focus on non-functional robustness and introduce RobuNFR, a benchmark designed to evaluate LLMs under real-world noisy contexts in question-answering (QA) tasks. Their results demonstrate that LLMs are highly sensitive to such noise, thereby extending robustness analysis beyond prompt-level perturbations. Mastropaolo et al.~\cite{mastropaolo2023robustness} conduct an empirical study on GitHub Copilot, showing that 46\% of Copilot’s outputs changed when given semantically equivalent paraphrases. This indicates a concerning lack of robustness in popular commercial models.

To improve robustness, researchers have proposed a variety of mitigation strategies. Agrawal et al.\cite{agrawal2025enhancing} introduce selective instruction augmentation and inference techniques, including self-denoising and representation alignment. They find that self-denoising achieves substantially higher performance gains than alternative strategies. Hu et al.\cite{hu2024prompt} propose two mitigation strategies for retrieval-augmented generation: robust prompt alignment, which maps perturbed prompts to semantically equivalent canonical forms to stabilize retrieval, and retrieval consistency filtering, which filters out prompts yielding inconsistent retrieval results across paraphrases. Wang et al.~\cite{wang2024resilience} explore instruction-level noise in instruction-tuned models and find that adversarial prompts can severely degrade task performance. They further propose robustness training using paraphrased instruction variants to improve model resilience.
Different from prior studies, we apply knowledge editing to enhance model robustness, without relying on additional ensemble strategies. Our method requires only a single example for intervention, offering a novel approach to improving the robustness of LLMs.

\vspace{-0.18cm}
\subsection{Knowledge Editing}
Knowledge editing has emerged as a promising direction for updating factual knowledge in LLMs without full retraining~\cite{zhong2023mquake,belrose2023leace,wei2023assessing,gupta2024model,hase2023does,hua2024propagation,lo2024large}, which can be categorized into three main paradigms~\cite{wang2024knowledge}. External memorization-based methods~\cite{madaan2022memory,zheng2023can,murty2022fixing,mitchell2022memory,li2022large,dong2022calibrating,huang2023transformer,hartvigsen2023aging,li2024consecutive,li2024swea} leverage an external memory to store new knowledge without modifying the pre-trained weights, thereby fully preserving the original knowledge encoded in the LLM. Global optimization-based methods~\cite{chen2020recall,sinitsin2020editable,lee2022plug,zhu2020modifying,ni2023forgetting,gu2024model,de2021editing,hase2023methods} aim to integrate new knowledge into pre-trained LLMs in a generalizable manner through optimization guided by the new information. These methods employ tailored strategies to constrain the impact on existing knowledge, distinguishing them from naive fine-tuning. The third category, local modification-based methods~\cite{dai2021knowledge,wu2023depn,li2024pmet}, is most relevant to our work. These methods aim to identify the parameters associated with specific knowledge in LLMs and selectively update them to incorporate new information related to the edit.

Meng et al.\cite{meng2022locating} propose ROME, a pioneering local editing method that identifies and updates key neuron activations within specific transformer layers responsible for storing factual associations. By analytically tracing causal dependencies, ROME directly modifies the internal representations of a model to reflect new factual information with high locality and minimal side effects. Building on this, Meng et al.\cite{meng2022mass} introduce MEMIT, a scalable framework that enables batch editing of multiple facts by optimizing a shared intervention across selected layers. Wang et al.~\cite{wang2024detoxifying} extend this line of work to the domain of safety, applying knowledge editing to mitigate toxic behaviors in LLMs by targeting offensive content and aligning representations with non-toxic alternatives. These methods underscore the potential of local interventions to precisely and efficiently influence LLM behavior. However, prior approaches have primarily focused on factual updates, whereas our work explores the novel application of local knowledge editing to improve robustness against prompt perturbations.

%% file: catalogue/conclusion.tex
In this paper, we presented \textbf{\appname}, a lightweight model editing framework to enhance the robustness of LLMs against natural language prompt perturbations. By leveraging a pair of original and perturbed prompts, \appname identifies robustness-sensitive layers through causal tracing and applies targeted parameter updates to align their internal representations. 
We conduct extensive experiments on the HumanEval and MBPP benchmarks using two representative code generation models. The results show that \appname substantially improves functional correctness on perturbed prompts, achieving a 63\% increase in Pass@1 accuracy, while maintaining performance on clean inputs within a $\pm$1\% margin.
We further assess generalization and show that \appname restores up to 30\% of robustness across perturbation types. Ablation studies confirmed the importance of each component, particularly layer localization. Our analysis of key layer distributions reveals that robustness-sensitive regions are concentrated in the middle-to-deep layers and vary with model architecture, offering actionable insights for future robustness-oriented interventions.
\appname provides a practical path toward making LLMs more reliable under real-world prompt variability, without requiring full retraining or architectural modifications. Future work will explore extending \appname to broader task domains and evaluating its effectiveness against more challenging perturbations.